\documentclass[prl,twocolumn]{revtex4} 
\usepackage[latin1]{inputenc}
\usepackage{dcolumn}
\usepackage[hypertex]{hyperref}

\newlength{\textlength}
\newlength{\overlinelength}

\newcommand{\abs}[1]{\left| #1\right|}
\newcommand{\VEV}[1]{\left\langle #1\right\rangle}

\begin{document}

\thispagestyle{empty}


\title{Higgs at the Tevatron in Extended Supersymmetric Models}

\author{Tim Stelzer, S\"oren Wiesenfeldt, and Scott Willenbrock}

\affiliation{Department of Physics, University of Illinois at
  Urbana-Champaign,
  \\
  1110 West Green Street, Urbana, IL 61801, USA}

\begin{abstract}
  \noindent
  Supersymmetric models with an additional singlet field offer the
  Higgs boson the possibility to decay to two pseudoscalars, $a$.  If
  the mass of these pseudoscalars is above the $b \bar{b}$ threshold,
  $a \to b \bar{b}$ is generically the dominant decay mode.  The decay
  $h \to a a \to b {\bar b} b {\bar b}$ may be seen above backgrounds
  at the Tevatron if the Higgs production cross section is enhanced
  relative to that of the standard model.
\end{abstract}


\maketitle

\emph{Introduction.}---
The Higgs boson has successfully resisted discovery as yet. Although
precision electroweak data, in combination with the direct top-quark
mass measurement at the Tevatron, hint at the existence of a light
scalar particle \cite{ewwg}, LEP has put a lower bound on the Higgs
mass within the standard model (SM), $M_h > 114.4$ GeV
\cite{Barate:2003sz}.

This bound has left some doubt as to whether the minimal
supersymmetric standard model (MSSM) is viable.  The issue is that
this model, which stabilizes the enormous hierarchy between the
electroweak and grand-unified or Planck scales, has a fine-tuning
problem unless the Higgs boson is somewhat lighter than the current
bound.

In addition, the MSSM suffers from the $\mu$-problem.  The
dimensionful parameter $\mu$ is required in order to give mass to
the Higgs boson and to communicate the electroweak symmetry breaking
between the two Higgs doublets.  Hence $\mu$ must be at the weak
scale, but na\"ively we would expect it to be of the order of the
grand-unified or Planck scale.

To solve the $\mu$-problem, several extensions have been proposed
where the $\mu$-parameter arises after an additional singlet field,
which does not interact with the MSSM matter and gauge fields,
acquires a vacuum expectation value (vev) \cite{Accomando:2006ga}.
The vevs of the Higgs doublets and the singlet are generically of the
same order.

The singlet field provides an additional scalar, a pseudoscalar, and
an accompanying Higgsino.  These mix with the neutral fields from
the two doublets, yielding five neutral Higgs bosons: three scalars
and two pseudoscalars.  In general, their masses are expected to be
comparable; on the other hand, these extended models possess
approximate U(1) symmetries, protecting the mass of one
pseudoscalar, $a$.  A light pseudoscalar is therefore natural,
allowing the decay $h \to a a$ (where $h$ is approximately SM-like)
with a branching ratio of nearly unity \cite{studies}. The
pseudoscalars then decay to fermion pairs, resulting in a
four-fermion final state, to which the LEP searches are less
sensitive \cite{Schael:2006cr}.  If the mass of $a$ is above the $b
\bar{b}$ threshold, the dominant final state is $b {\bar b} b {\bar
b}$ \cite{cm}.

In this paper we propose exploring such a scenario at the Tevatron.
The Higgs boson is dominantly produced singly and subsequently
decays via $h \to a a \to b {\bar b} b {\bar b}$.  We calculate the
backgrounds to this signal using the multi-purpose code MadEvent
\cite{madgraph}.  It is usually assumed, either explicitly or
tacitly, that this background overwhelms the signal, and we confirm
that this is the case.  However, we find that if the signal is
sufficiently enhanced, then it emerges from the background. This
opens the question of whether there exists models with enhanced
Higgs production and with a significant branching ratio for the
above decay mode.

We do not restrict our study to one particular model beyond the MSSM
but consider the general case, where $M_h$ varies between 110 and 150
GeV.  This approach is motivated by the fact that these extended
models include a large region in parameter space with $M_h$ in this
range and with $M_a$ between zero and 200 GeV \cite{Barger:2006dh}.

In this mass region, the SM Higgs production cross section at the
Tevatron is less than 1 pb, via $gg\to h$; however, in the MSSM the
cross section is much larger for large $\tan\beta$, with both $g g \to
h$ and $b \bar{b} \to h$ contributing \cite{Hahn:2006my}.  It is an
open question whether there exist extensions of the MSSM which
maintain this large production rate, while at the same time yielding a
significant branching ratio for $h \to a a \to b {\bar b} b {\bar b}$.

\emph{Background.}---
Let us consider the general case of a scalar particle, $h$, which
almost exclusively decays to two lighter pseudoscalars (or scalars),
$a$, followed by the decay to $b$ quarks.

The dominant background is due to QCD multijet production, with
varying combinations of true $b$ tags and mistagged jets.  As we will
see, we must require at least three $b$ tags to have a reasonable
signal-to-background ratio, so we have to consider the backgrounds
($j=u,d,s,c,g$)
\begin{itemize}
  \addtolength{\itemsep}{-7pt}
\item $p \bar{p} \to b \bar{b} b \bar{b}$;
\item $p \bar{p} \to b \bar{b} b j$;
\item $p \bar{p} \to b \bar{b} j j$, where one jet is mistagged;
\item $p \bar{p} \to b j j j$, where two jets are mistagged;
\item $p \bar{p} \to j j j j$, where three jets are mistagged.
\end{itemize}

The CDF and D0 collaborations have performed searches for neutral
Higgs bosons produced in association with bottom quarks, followed by
$h \to b \bar{b}$, using a secondary-vertex trigger \cite{exp}. Guided
by their analyses, we chose the cuts listed in Table~\ref{tb:cuts}.
The requirement on the minimum invariant mass of any two jets may not
be necessary, but it eliminates many background events and therefore
makes the event generation more efficient.

\begin{table}
  \caption{Parameters and cuts.
    \label{tb:cuts}}
  \begin{ruledtabular}
    \begin{tabular}{ll}
      {\bfseries Parameters} & \cr
      renormalization scale & $\VEV{p_T}$ \cr
      factorization scale & $\VEV{p_T}$ \cr
      PDF & CTEQ6L \cr
      $b$ mass & $m_b = 0 $ \cr
      \hline
      {\bfseries Cuts} & \cr
      rapidity & $\abs{\eta} < 2.0$ \cr
      separation & $\Delta R > 0.4$ \cr
      jet 1 & $p_T>20$ GeV \cr
      jets 2--4 & $p_T>15$ GeV \cr
      invariant mass of two jets & $m_{jj} > 10$ GeV \cr
      \hline
      {\bfseries Tagging efficiencies} & \cr
      $b$ tag & 50\% \cr
      mistag of $c$ & 10\% \cr
      mistag of light quark or gluon & \phantom{0}1\%
    \end{tabular}
  \end{ruledtabular}
\end{table}

The different background processes sum to an enormous background of
380 nb prior to $b$ tagging.  In order to extract the signal, we must
require that three or more jets are tagged.  In reality, the tagging
efficiency is a $p_T$ and $\eta$ dependent function.  For simplicity,
and to allow others to easily reproduce our results, we approximate
the tagging efficiency and the mistag rates by the constant values
listed in Table~\ref{tb:cuts}.  This overestimates the actually
capabilities of the detectors, but is sufficient for a crude analysis.

Tagging three or more jets, the background drops dramatically to 63
pb.  Table~\ref{tb:background-xs} lists the cross sections of the
various processes, categorized by the number of $b$ and $c$ jets
present. We see that $b \bar{b} j j$ with one mistagged light jet
makes up about half of the background, followed by $b\bar bbj$ and
$b\bar bb\bar b$.  The largest backgrounds with mistagged $c$ jets are
$b\bar bcj$ and $b\bar bc\bar c$, but they are relatively small.

\begin{table}
  \caption{The cross sections (pb) of the various background
    processes $p \bar{p} \to j j j j$ after the cuts and
   tagging efficiencies of Table~\ref{tb:cuts}.  The cross sections
   are organized by the number of $b$ and $c$ jets in the event.
    \label{tb:background-xs}}
  \begin{ruledtabular}
    \begin{tabular}{l|d|d|d|d|d|d}
      \multicolumn{1}{l|}{}
      & \multicolumn{1}{c|}{total} & \multicolumn{1}{c|}{$n_c=0$} &
      \multicolumn{1}{c|}{$n_c=1$} & \multicolumn{1}{c|}{$n_c=2$} &
      \multicolumn{1}{c|}{$n_c=3$} & \multicolumn{1}{c}{$n_c=4$} \cr
      \hline
      total & 63 & 54 & 4 & 5 & 0.2 & 0.1 \cr
      \hline
      $n_b=0$ & 3 & 0.8 & 0.2 & 1 & 0.2 & 0.1 \cr
      \hline
      $n_b=1$ & 1 & 0.5 & 0.05 & 0.5 & 0 \cr
      \cline{1-6}
      $n_b=2$ & 40 & 33 & 4 & 3 \cr
      \cline{1-5}
      $n_b=3$ & 10 & 10 & 0.1 \cr
      \cline{1-4}
      $n_b=4$ & 9 & 9 \cr
    \end{tabular}
  \end{ruledtabular}
\end{table}

Let us now consider different windows in the
$\left(M_h,M_a\right)$-plane, where we choose the masses of $h$ and
$a$ to be $M_h = 110,\ 130,\ 150$ GeV and $M_a = 20,\ 40,\ 60$ GeV.
The jets are paired such that their invariant masses are as close as
possible. The windows have a size of 30 GeV for the invariant $b
\bar{b}$ and $b \bar{b} b \bar{b}$ masses, again guided by
Ref.~\cite{exp}. The results are shown in
Table~\ref{tb:background-mh-ma}; the background is between 10 and 15
pb for all masses considered.

\begin{table}
  \caption{Background cross sections (pb) for different choices of
    $M_h$ and $M_a$ with window sizes of 30 GeV.
    \label{tb:background-mh-ma}}
  \begin{ruledtabular}
    \begin{tabular}{c|c|c|c}
      & $M_a = 20$ GeV & $M_a = 40$ GeV & $M_a = 60$ GeV \cr
      \hline
      $M_h = 110$ GeV & 15 & 14 & 12 \cr
      \hline
      $M_h = 130$ GeV & 15 & 15 & 13 \cr
      \hline
      $M_h = 150$ GeV & 11 & 11 & 11 \cr
    \end{tabular}
  \end{ruledtabular}
\end{table}

\emph{Signal.}---
The signal events have to pass the same cuts as the background
processes (see Table~\ref{tb:cuts}).  Table~\ref{tb:signal-eff} shows
the product of the acceptance and tagging efficiency for the different
choices of $M_h$ and $M_a$.

\begin{table}
  \caption{Acceptance $\times$ tagging efficiency of the signal for
    different choices of $M_h$ and $M_a$.
    \label{tb:signal-eff}}
  \begin{ruledtabular}
    \begin{tabular}{c|c|c|c}
      & $M_a = 20$ GeV & $M_a = 40$ GeV & $M_a = 60$ GeV \cr
      \hline
      $M_h = 110$ GeV  & 0.04 & 0.04 & --- \cr
      \hline
      $M_h = 130$ GeV  & 0.06 & 0.05 & 0.14 \cr
      \hline
      $M_h = 150$ GeV  & 0.09 & 0.08 & 0.12 \cr
    \end{tabular}
  \end{ruledtabular}
\end{table}

For a discovery of $h$, the ratio $S/\sqrt{B}$, where $S$ and $B$ are
the signal and background, must be at least five.  We use
2~$\text{fb}^{-1}$ of integrated luminosity to derive the minimum
signal cross section for a discovery of $h$.  We assume that all
signal events pass the mass reconstruction constraints, which is a
good approximation.  We use an ideal branching ratio for $h \to a a
\to b {\bar b} b {\bar b}$ of 100\%; the minimum signal cross section
is increased by a factor of $1/\mathit{BR}$ for other branching
ratios.  The results are given in Table~\ref{tb:signal-limit}.

If one tags all four jets, the tagging efficiency for the signal
drops by a factor of 5, due in part to combinatorics.  Looking at
Table~\ref{tb:background-xs}, one finds that the only significant
background with all four jets tagged is $b\bar bb\bar b$, and this
also drops by the same factor of 5.  Since this background 1/7 of
the total background with three or more tags, the overall gain in
signal significance with four tags is a modest factor of
$\sqrt{7/5}\approx 1.2$.

\begin{table}
  \caption{Discovery cross section for the signal (pb) with
    2~$\text{fb}^{-1}$ of data if all signal events pass the mass
    reconstruction constraints, assuming a branching ratio for
    $h \to a a \to b {\bar b} b {\bar b}$ of 100\%.
    \label{tb:signal-limit}}
  \begin{ruledtabular}
    \begin{tabular}{c|c|c|c}
      & $M_a = 20$ GeV & $M_a = 40$ GeV & $M_a = 60$ GeV \cr
      \hline
      $M_h = 110$ GeV  & 12 & 11 & --- \cr
      \hline
      $M_h = 130$ GeV  & \phantom{0}7 & \phantom{0}9 & \phantom{0}3
      \cr
      \hline
      $M_h = 150$ GeV  & \phantom{0}4 & \phantom{0}5 & \phantom{0}3
      \cr
    \end{tabular}
  \end{ruledtabular}
\end{table}

\emph{Discussion.}---
The minimum cross section required for discovery is an order of
magnitude greater than the SM Higgs production cross section,
confirming the belief that the backgrounds overwhelm the signal in
this case. Increasing the integrated luminosity to 8~$\text{fb}^{-1}$
decreases the minimum cross section by a factor of two, still not
enough to discover a SM-like Higgs in this decay mode.  However, if
there exist models in which the Higgs production cross section is
enhanced by an order of magnitude, while still maintaining a
significant branching ratio for $h\to aa\to b\bar bb\bar b$, then it
appears possible to discover such a Higgs at the Tevatron.  This is an
open question in extensions of the MSSM.

\emph{Acknowledgments.}---
We are grateful for conversations with P.~Bechtle, P.~Fox, T.~Junk,
T.~Liss and K.~Pitts.  S.~W.~thanks the Aspen Center for Physics for
hospitality.  This work was supported in part by the U.~S.~Department
of Energy under contract No.~DE-FG02-91ER40677.



\begin{thebibliography}{19}
\bibitem{cm}%
  Scenarios where other decay channels become dominant have been
  studied in Refs.~\cite{studies} as well.
\end{thebibliography}
\end{document}